\documentclass[conference,compsoc]{IEEEtran}
\usepackage{graphicx}

\ifCLASSOPTIONcompsoc
    \usepackage[nocompress]{cite}
\else
  \usepackage{cite}
\fi

\begin{document}

\title{A NIS Directive compliant Cybersecurity Maturity Assessment Framework}

\author{\IEEEauthorblockN{George Drivas}
\IEEEauthorblockA{Department of Digital Systems \\ University of Piraeus \\ Piraeus, Greece \& \\
National Cyber Security Authority of Greece}
\and
\IEEEauthorblockN{Argyro Chatzopoulou}
\IEEEauthorblockA{APIROPLUS Solutions \\ Limassol, Cyprus}
\and
\IEEEauthorblockN{Leandros Maglaras}
\IEEEauthorblockA{Faculty of Computing, Engineering and Media\\ De Montfort University \& \\
National Cyber Security Authority of Greece
}
\and
\IEEEauthorblockN{Costas Lambrinoudakis}
\IEEEauthorblockA{Department of Digital Systems \\ University of Piraeus \\ Piraeus, Greece}
\and
\IEEEauthorblockN{Allan Cook \\ and Helge Janicke}
\IEEEauthorblockA{Faculty of Computing, Engineering and Media\\ De Montfort University
}
}

\maketitle

\begin{abstract}
The NIS Directive introduces obligations for the security of the network and information systems of operators of essential services and of digital service providers and require from the national competent authorities to assess their compliance to these obligations. This paper describes a novel cybersecurity maturity assessment framework (CMAF) that is tailored to the NIS Directive requirements and can be used either as a self assessment tool from critical national infrastructures either as an audit tool from the National Competent Authorities for cybersecurity. 
\end{abstract}

\IEEEpeerreviewmaketitle

\section{Introduction}
Cyber attacks could contribute towards the collapse of a state if they initiate or prolong the failure of Critical National Infrastructures (CNI). Nations are becoming reliant on the cyber domain to provide services that keep a nation running: power grids, water supplies, communications,
transportation and finance are all increasingly becoming cyber dependant. 

The NIS Directive (Directive (EU) 2016/1148) (European Union, 2016) has certain obligations that each member state should follow, with a major goal of enhancing cybersecurity posture across the EU. Having to cope with the obligations of the NIS directive and to meet strict deadlines, Greece has taken some steps forward \cite{maglaras2018nis}. 
The directive and a relevant national Law (Greek National Law 4577/2018)(Greek Government, 2018) introduce obligations for the Security Of The Network And Information Systems Of Operators Of Essential Services (OES) and the Security Of The Network And Information Systems Of Digital Service Providers (DSP) and require from the National Competent Authorities (NCA) to assess the compliance of operators of essential services and digital service providers with these obligations. 

More specifically, the Greek National Law 4577/2018 (Greek Government, 2018) states that “the National Cybersecurity Authority (NCSA), acting as the NCA for cybersecurity, in collaboration with the relevant CSIRTs and other organizations and entities as appropriate assesses the technical and organizational measures implemented by OES, in order to manage risks related to the security of network and information systems used in their activities, regarding their suitability and their proportionality". Additionally, NCSA "assesses the suitability of the measures implemented by DSP for the avoidance and the minimization of the impact of incidents affecting the security of network and information systems used for the provision of the basic services, aiming to assure their business continuity".

Moreover, the objectives of the NCA across the EU reach further than just the collection of evidence send by the assessed entities and include the vision to reach a common level of cybersecurity posture. In order to effectively meet this vision, an initial step for NCSA was to conduct an assessment of the current cybersecurity posture of public sectors' main ICT services, using a structured questionnaire. This assessment revealed inconsistencies and major misalignment among different entities \cite{drivas2019cybersecurity}. As a result, the below additional targets are proposed:
\begin{itemize}
\item Standardization of the collected feedback
\item Assignment of a specific level of security, based in the implemented controls per category
\item Analysis of the outputs and extraction of relevant statistical information regarding the level achieved per industry, category and service
\item Implementation of comparisons between subsequent assessments, in order to monitor progress
\item Extraction of possible correlations or contrasts between the information security posture among stakeholders
\item Conduction of further analysis and definition of best practices
\end{itemize}

The minimum security requirements that OES and DSP have to comply with, have been defined in Decision 1027 published in 3739, B, 08.10.2019 Official Gazette of the Greek Government (Greek Government, 2019).
This set of requirements, covering areas like Risk Management, Access Control, Physical and Environmental Controls etc, is generic in phrasing and does not provide specifications regarding the requirements implementation. For example, for the area of physical and environmental security, the requirement is that the installations of data centers and information processing facilities shall be protected against physical or environmental risk through suitable and relevant policies and measures based on a risk management strategy (Greek Government, 2019).
This phrasing, although mandated by the fact that the entities required to comply with these requirements have a great variety in terms of business operation, size, security posture and technical and organizational capability, is difficult to be monitored effectively by the NCA in order to achieve the objectives mentioned above.

What was needed in order to facilitate the fulfillment of the NCSA’s objectives, especially the measurable ones, was a tool to standardize the possible maturity levels of the organizations. For this purpose a specific assessment framework, the Cybersecurity Maturity Assessment Framework (CMAF), was designed and tested for implementation.
The CMAF could help identify the strengths and weaknesses of an organization’s processes and examine how closely these processes comply to related identified best practices or guidelines.

The assessment framework consists of the security controls against which the organization’s processes are appraised and the scale, based on which the rating of compliance of the organization’s processes is evaluated. Based on the proposed targets mentioned above the assessment model should incorporate the following characteristics:

\begin{itemize}
\item Cover the full extent of the security requirements complying to the NIS directive obligations
\item Be able to be used as a self-assessment tool
\item Be able to be used as a basis for an independent assessment
\item Provide clear results regarding the security posture of the organizations
\item Be able to be used as a benchmarking tool per industry, type of organization and area of operation
\item Be able to be used as a guide for security requirements implementation by the organizations
\item Be measurable
\item Be easily extractable
\end{itemize}

\subsection{Design of the framework}
In order to design the CMAF, a combination of literature review regarding security requirements and a review of existing frameworks (related to cybersecurity or other well established areas) was conducted. 
At the time of the conduction of this review, there was only a limited number of established frameworks in the field, although during the past months, some more have been introduced. The literature review regarding security requirements included 16 basic documents. These documents were published by organizations like ENISA, ISO, CIS, European Union, NIST, ISACA and others. The review regarding existing frameworks included frameworks or models from organizations like: CMMI, CIS, ENISA, Department of Homeland Security – USA, Citigroup, U.S.Department of Energy and others.

\subsection{The maturity scale}

\begin{figure*}[t]
\includegraphics[width=\textwidth]{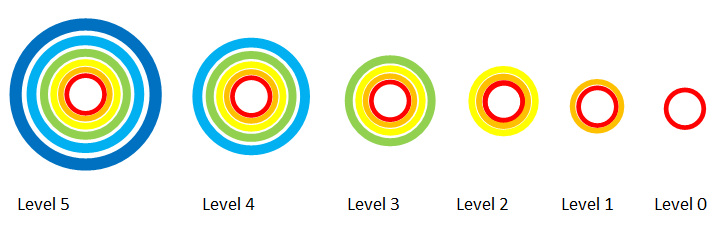}
\caption{Maturity levels representation}\label{figure1}
\end{figure*}

After the review of the existing frameworks, it was decided that the CMAF would be based in a 6-levels maturity scale (Figure \ref{figure1}). For each level of the maturity scale, a different seal was selected. The seals represent a series of concentric cycles. The lowest possible score being represented by one cycle and the highest by six. The color of each cycle has been selected from the PH scale – red being the lowest and blue the highest.

\begin{itemize}
    \item  {\bf Maturity Level 5: Efficient - Optimized}
 
  The organization has implemented methods for the continuous improvement of the implemented controls and the security posture of the organization. A full risk-based approach is followed and a cost-benefit balance is applied. The necessary controls are implemented, measured and controlled at the described level. 
\item  {\bf Maturity Level 4: Effective - Quantitatively Managed}

The organization has set relevant objectives. The objectives (were possible S.M.A.R.T.) are being monitored, measured, analyzed and evaluated. The necessary controls are implemented, measured and controlled at the described level.

\item {\bf Maturity Level 3: Advanced - Defined}

There is a standardized method regarding the fulfillment of the requirement. The necessary controls are implemented, measured and controlled at the described level.

\item {\bf Maturity Level 2: Basic - Managed}

There is a concrete plan regarding the fulfillment of the requirement. The necessary controls are implemented, but are partially measured and partially controlled.

\item {\bf Maturity Level 1: Initial - Reactive}

The organization has started implemented the requirement, but the extent of the implementation is partial or reactive.

\item {\bf  Maturity Level 0: Incomplete - Not existing}

The requirement under examination is not implemented or it is implemented only partially or ad hoc.

\end{itemize}

\subsection{The requirements}

The basis of the CMAF and thus the requirements of the framework, are those that have been published under the sections “Common Security Policy” and "Baseline Security Requirements", in Ministerial Decree 1027 - 3739, B, 08.10.2019 Official Gazette of the Greek Government (Greek Government, 2019). 
These requirements are grouped under the following categories:
\begin{itemize}
\item	Information Security Policy
\item	Business Environment
\item	Asset Management
\item	Risk Assessment
\item	Risk Management Strategy
\item	Supply Chain Risk Management
 \item Self-Assessment and Improvement
\item  Policies, Processes and Procedures for the protection of essential services
\item	Identity Management and Access Control
\item	Physical and Environmental Security
\item	Systems and Applications Security
\item	Data Security
\item 	Backups
\item Security Technologies
\item	Systems Testing
\item	Change Management
\item Awareness and Training
\item	Threat Detection
\item Incident Management
\item	Business Continuity
\item	Disaster Recovery
\end{itemize}
For each of these categories the applicable controls have been recognized, analyzed and assigned to the appropriate level, as described in the maturity scale above. 

\subsection{The framework structure}

The structure of the resulting framework is the following:
\begin{itemize}

\item A. IDENTIFICATION	

Requirement A1. Business Environment	

Requirement A2. Asset Management	

Requirement A3. Risk Assessment	

Requirement A4. Risk Management Strategy	

Requirement A5. Supply Chain Risk Management	

Requirement A6. Self-Assessment – Improvement	

\item B. PROTECTION	

Requirement  B7. Policies, Processes and 
Procedures for the protection of essential services.

Sub - Requirement B7.1 : Information Security Policy, Processes and Procedures

Sub - Requirement B7.2 : Communication and acceptance

Requirement B8. Identity Management and Access Control	

Sub - Requirement B8.1 Asset Management

Sub - Requirement B8.2. Access control for privileged accounts

Sub - Requirement B8.3. Management of equipment for administrative purposes

Sub - Requirement B8.4. Access Control

Sub - Requirement B8.5. Authentication mechanisms

Requirement B.9. Physical and Environmental security	

Requirement B.10. Systems and Applications security	

Sub - Requirement  B.10.1.: Systems  Security

Sub - Requirement B.10.2.: Application Security

Sub - Requirement B.10.3.: Security in 
Application Development

Requirement B.11. Data Security	

Sub - Requirement B.11.1.: Encryption

Sub - Requirement B.11.2.: Data Classification

Requirement B.12. Backups

Requirement B.13. Security Technologies	

Sub - Requirement B.13.1. : Traffic filtering

Sub - Requirement B.13.2.: Segregation of 
systems

Sub - Requirement B.13.3. : Malware protection

Requirement B.14. Systems Testing	

Sub - Requirement B.14.1.: Security Assessments

Sub - Requirement B.14.2.: Compliance Checking

Requirement B.15. Change Management	

Requirement B.16. Awareness and Training	

\item C. DEFENSE	

Requirement C.17. Threat Detection	

Requirement C.18. Incident Management

Requirement C.19. Business Continuity	

Requirement C.20 Disaster Recovery	
\end{itemize}

\subsection {Framework validation}

Since the framework was primarily based on literature review, it was decided that before it’s release, it should be validated through an implementation pilot project. 
The pilot project should include organizations from different sectors, of different sizes and with different security postures.
To achieve this, the following pilots were selected for validation purposes:
\begin{itemize}
    \item 	A mid-sized OES from healthcare sector, that is expected to have a low level of maturity
\item	A mid-sized OES from Digital Infrastructures sector, that is expected to have a medium level of maturity
\item	A large-sized OES from air-transport sector, that is expected to have a high level of maturity
\end{itemize}

The pilot assessments were carried out by an experts team consisting of: One expert from the project team, involved in the development of the CMAF with a deep knowledge of the framework itself and with high security control auditing skills, and two experts from the NCSA, tasked with the monitoring of the framework implementation process and the review of the related outcomes. 

\begin{figure}[t]
\includegraphics[width=0.5\textwidth]{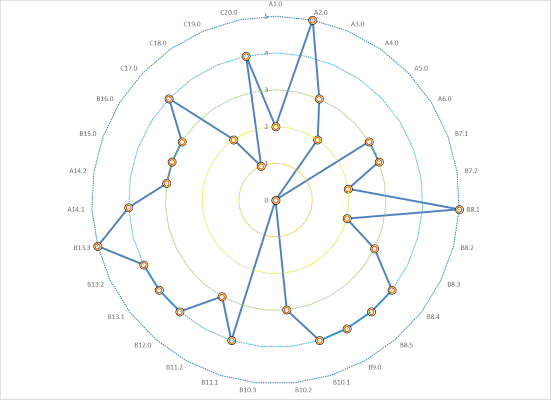}
\caption{Graphical representation of the security assessment}\label{figure2}
\end{figure}

The assessments were carried out via the following methods: Table-top assessment, Interview and On site visit. 
The results of the assessments and the improvements proposed, were gathered, analyzed and incorporated in the final version of the model. In Figure \ref{figure2} the graphical representation of the assessment result of a fictitious organization, that is produced from the model implementation, is presented.

\section{Related Work}
Authors in \cite{rea2017maturity} describe a cybersecurity capability maturity model as a means by which an organisation can assess its current level of maturity of its practices.  They provide a comparative study of cybersecurity capability maturity models that builds on a previous review \cite{rea2017comparative}.  The research presents an assessment of the differences, advantages and disadvantages of a systematic review of published studies from 2012 to 2017.The method was based on a modified taxonomy of software improvement environments across five categories proposed by Halvorson and Conradi \cite{halvorsen2001taxonomy}:

\begin{enumerate}
\item	General: The broad attributes of the improvement environment.
\item	Process: Describing how the environment is used.
\item	Organisation: The features that articulate the relationship between the organisation and the environment in which it is used.
\item	Quality: The indicators used to determine quality in the environment.
\item	Result: A statement of the required outcomes, associated costs, and the methods used for evaluation.
\end{enumerate}

Rea-Guaman et al \cite{rea2017comparative} reduce the categories to three, arguing that quality and result do not support the comparison of cybersecurity capability maturity models.

The research considers the Systems Security Engineering Capability Maturity Model (SSE-CMM), Cybersecurity Capability Maturity Model (C2M2) \cite{christopher2014cybersecurity}, Community Cyber Security Maturity Model (CCSMM) \cite{white2011community}, National Initiative for Cybersecurity Education – Capability Maturity Model (NICE) \cite{curtis2015cybersecurity}.

The analysis concludes that all four models require a degree of customisation, and that SSE-CMM is the most established of those reviewed.   The paper assessed C2M2 as the only model focused on cybersecurity.  It went on to state that C2M2 and CCSMM are designed to be implemented in conjunction with the NIST framework and that recent updates to the models had not been identified.  It further concluded that only SSE-CMM and C2M2 provided detailed considerations of risk.   The research did not present a model to address the issues identified within those reviewed.

Sabillon \cite{sabillon2017comprehensive} reviews the best practices and methodologies in cybersecurity assurance and audit and presents the Cyber Security Audit Model (CSAM) for use by organisations and nation states to validate audit, preventative, forensic and detective controls. CSAM comprises 18 domains, with one limited to nation states and the remaining 17 applicable to organisations in general.  All domains have at least one sub-domain, controls, checklists, sub-controls, and scorecard.  The authors compare CSAM to the NIST Cyber Security Framework (CSF) version 1.1 and the Audit First Methodology \cite{donaldson2015enterprise}, highlighting the differences.  The work describes the CSAM model and states it was tested, implemented and validated in a Canadian higher education institution, although no results are presented, and its efficacy is not evidenced. 

The research does not indicate whether CSAM requires specific expertise for its use, or whether self-assessment is feasible.  Neither does it present whether the model provides any actionable outcomes following its use.

Akinsanya \cite{akinsanya2019current} performed a literature review of cyber security models for healthcare organisations adopting cloud computing.  The analysis considers the following:
\begin{itemize}	
\item Information Security Focus Area Maturity Model (ISFAM)\cite{van2010design}
\item	Cloud Security Capability Maturity Model (CSCMM)
\item	UK National Health Service (NHS) National Infrastructure Maturity Model (NIMM)\cite{savidas2009your}
\item	Health Information Network (HIN)
Capability Maturity Model
\end{itemize}

The analysis found that ISFAM was targeted on small to medium enterprises with a focus of use within software development environments and could not demonstrate a capacity to integrate emerging technologies such as cloud computing.  CSCMM was identified as able to support a range of organisational sizes but deemed too technically complex to implement in healthcare.  The NHS NIMM was assessed as relevant to the cyber security assessment of a healthcare organisation.  It was considered to provide an assessment of an organisation’s cyber security maturity and provided a basis to define which steps were required to achieve a greater level of maturity. It demonstrated a capacity to support platform independence but did not show an ability to accommodate the characteristics of cloud computing or its resulting threats.  The researchers judged that the HIN Capability Maturity Model has similar characteristics and constraints as the NHS NIMM.  

The researchers concluded that maturity models provide a compliance model that could not support the complexity of the emerging cyber environment, particularly for healthcare organisations adopting cloud technologies.  The conclusions highlight three specific areas of concern and requirements for further work: 
\begin{itemize}
\item	Cyber security maturity models should focus on more than standards compliance.
\item	Any new measures of maturity introduced should be provided with adequate metrics to make them meaningful.
\item	The model should be extensible to accommodate the dynamic nature of the cyber security threat landscape.
\end{itemize}

The research focuses specifically on healthcare and the adoption of cloud computing and does not consider adoption beyond this industry or technology.

Miron and Muita \cite{miron2014cybersecurity} examine cybersecurity maturity models to identify the standards and controls available to providers of critical national infrastructures (CNI).  The research considers the cyber threats to CNI and the impact of a loss of such services.    The authors document nine cybersecurity capability maturity models assessed as applicable to CNI.  These are presented based on their applicability to either specific CNI sub-sectors or their general cyber security focus.  The review concludes that, of the models considered, none are designed to address the scenario of a CNI operator with a multiplicity of interdependent systems.  Instead, the authors propose, the models are described at a high level and focus on CNI or industry sub-sectors and present the need for a model to support municipal governments.

Adler \cite{adler2013dynamic} states that capability maturity models are inherently static and diagnostic, in that they identify maturity gaps but are not directly actionable.  The research proposes a methodology that follows three stages:

\begin{enumerate}
\item	Model: Captures the organisation’s current-state cybersecurity maturity levels, formulates a maturity improvement plan, and identifies influences factors that will shape the organisation’s cybersecurity posture.
\item	Simulate: Produces dynamic simulations of scenarios to determine how particular cybersecurity situations could evolve within the organisation, potential interventions, and likely outcomes and impacts.
\item	Analyse: Assesses the projected mitigation costs and risk reduction benefits, comparing outcomes across alternate plans and scenarios.
\end{enumerate}

The stages are illustrated through an analysis based upon extension to the Electric Sub-sector Cybersecurity Capability Maturity Model (ES-C2M2) \cite{stevens2014electricity}.  ES-C2M2 is a lightweight adaptation of the Resilience Maturity Model \cite{caralli2010cert}.  This extension describes that the requirement for any process improvement elements within the maturity model should provide explicit guidance for improving performance levels towards a desired end-state.

The research describes a set of decision support tools implemented in a proprietary software package but does not provide data or results from experimental use of the proposed approach.

Le and Hoang in \cite{le2016can} investigate existing cybersecurity maturity models to examine their strengths and weaknesses.  They provide a comparison of 12 models mapped against five levels of maturity proposed by Humphrey (1989), concluding that models require relevant quantitative metrics for measurable and actionable assessment.  No examples of such metrics are provided.

Almuhammadi and Alsaleh in \cite{almuhammadi2017information} review the NIST Cyber Security Framework (CSF) for critical infrastructures in order to assess how it can be applied to cybersecurity maturity models and to determine if any gaps exist.  The authors propose that one of the key benefits of a cybersecurity capability maturity model is that it provides a structure to allow stakeholders to reach a consensus and set agreed priorities. The authors state that the NIST CSF does not provide organisations with a mechanism to measure the progress of a NIST implementation, or the maturity levels and information security processes’ capabilities.  They assess that the framework focuses on high-level information security requirements and is applicable for the development of information security programmes and policies.  They contrast this to other frameworks such as COBIT, ISO 27001, and the ISF Standard of Good Practice (SoGP) for Information Security.  They argue that these focus on information security technical and functional controls, and that they are applicable for developing checklists and conducting compliance/audit assessments. 
The research proposes an information security maturity model that is able to measure the implementation progress of a security programme over time and assesses the reliability of the IT services that underpin a business.  No examples of such measures are provided, and no experimental data is presented.

\section{Discussion}

The National Competent Authorities for cybersecurity, especially those who need to comply with the NIS directive, could use the proposed maturity assessment framework in order to request from the applicable organizations the implementation of self-assessments based on the framework and collect the results. The authorities can review the collected responses, analyze the data and produce valuable conclusions. Also, comments regarding possible improvements can be collected by the authorities from the applicable organizations as well as from the dedicated staff dealing with the assessment of the results.
Adapted version of the CMAF or other similar models incorporating specific requirements for cloud services and OT/ IoT environments could also be used. Based on the findings of the cybersecurity assessment of the Critical National Infrastructures, additional cyber security controls may be applied. Those security controls can be classified according to legal, technical, organizational, capacity building, and cooperation
aspects
\section{Conclusions}
As Critical National Infrastructures are becoming more vulnerable to cyber attacks, their protection becomes a significant issue for any organization, as well as a nation. Similar to other information technology (IT) processes, cybersecurity often follows a lifecycle model of prediction, protection, detection, and reaction. Moreover, an assessment is an activity that helps identify the strengths and weaknesses of an organization’s processes and examine how closely these processes relate to identified best practices and guidelines. In order to help in the evaluation of the cybersecurity posture of CNI, a novel cybersecurity maturity assessment framework, the CMAF, is presented in this paper. The proposed framework consists of 20 security categories, 6 maturity levels and and can be used both as a self assessment and as an external audit tool, facilitating organisations to perform a gap analysis and receive graphical representation of their security posture. Information that will be collected from the framework can be used, after proper aggregation and anonymisation processes, from National Competent Authorities in order to identify  common security gaps and prioritise future security programmes and funding actions on a national or European level.

\ifCLASSOPTIONcompsoc
  \section*{Acknowledgments}
  We thankfully acknowledge the support of the CONCORDIA H2020 (GA no. 830927) EU project.
\else
  \section*{Acknowledgment}
\fi
\bibliographystyle{IEEEtran}
\bibliography{george}	

\end{document}